\documentclass[a4paper,fleqn,usenatbib]{mnras}

\usepackage{newtxtext,newtxmath}

\usepackage[T1]{fontenc}
\usepackage{ae,aecompl}

\usepackage{graphicx}	
\usepackage{amsmath}	
\usepackage{amssymb}	
\usepackage{hyperref}       
\usepackage{newtxtext,newtxmath}
\usepackage{multirow}
\usepackage{amsfonts}
\usepackage{adjustbox}
\usepackage{placeins}

\def\CN2{\mbox{$C_N^2 $}}
\def\CT2{\mbox{$C_T^2$}}
\def\tauO{\mbox{$\tau_{0} $}}
\def\thetaO{\mbox{$\theta_{0} $}}

\title[Filtering methods to enhance forecast performances]{Filtering techniques to enhance optical turbulence forecast performances at short time scales}

\author[E.Masciadri et al.]{
E. Masciadri,$^{1}$\thanks{E-mail: elena.masciadri@inaf.it}
G. Martelloni,$^{1}$
A. Turchi$^{1}$
\\
$^{1}$INAF Osservatorio Astrofisico di Arcetri, Largo Enrico Fermi 5, I-501 25 Florence, Italy}

\date{Accepted XXX. Received YYY; in original form ZZZ}

\pubyear{2019}

\begin{document}
\label{firstpage}
\pagerange{\pageref{firstpage}--\pageref{lastpage}}
\maketitle

\begin{abstract}
The efficiency of the management of top-class ground-based astronomical facilities supported by Adaptive Optics (AO) relies on our ability to forecast the optical turbulence (OT) and a set of relevant atmospheric parameters. Indeed, in spite of the fact that the AO is able to achieve, at present, excellent levels of wavefront corrections (a Strehl Ratio up to 90\% in H band), its performances strongly depend on the atmospheric conditions. Knowing in advance the atmospheric turbulence conditions allows an optimization of the AO use. It has already been proven that it is possible to provide reliable forecasts of the optical turbulence ($\CN2$ profiles and integrated astroclimatic parameters such as seeing, isoplanantic angle, wavefront coherence time, ...) for the next night. In this paper we prove that it is possible to improve the forecast performances on shorter time scales (order of one or two hours) with consistent gains (order of 2 to 8) employing filtering techniques that make use of real-time measurements. This has permitted us to achieve forecasts accuracies never obtained before and reach a fundamental milestone for the astronomical applications. The time scale of one or two hours is the most critical one for an efficient management of the ground-based telescopes supported by AO. We implemented this method in the operational forecast system of the Large Binocular Telescope, named ALTA Center that is, at our knowledge, the first operational system providing  forecasts of turbulence and atmospheric parameters at short time scales to support science operations.
\end{abstract}



%
%
\section{Introduction}
\label{intro}

In spite of the fact that the Adaptive Optics (AO) is able to achieve, at present, excellent levels of correction of the perturbed wavefront (Strehl Ratio up to 90\% in H band on high contrast imaging SCAO\footnote{SCAO states for Single Conjugated Adaptive Optics} systems), the AO performances are strongly dependent on the atmospheric conditions. A couple of examples are emblematic in this respect. Performances of the best SCAO systems for 8-10m class telescopes can achieve a Strehl Ratio (SR) in H band of 90\% with a seeing of the order of 0.4" but the SR can drastically decreases to 20\% if the seeing is of the order of 1.2". Looking at the problem from a different point of view, if the seeing improves from 1" to 0.6", the limit magnitude of the AO guide stars with which we obtain a SR of 30\% can move from 13 mag to 15 mag for the same instrument. Such a better seeing strongly increases the sky coverage and opens new observational windows and new perspectives in terms of scientific programs. This gain in magnitude should permit, for example, to increase by a factor 10  the number of accessible AGNs from the ground (from the order of 10 to the order of 100).

The efficiency of modern ground-based astronomy, particularly if supported by Adaptive Optics (AO) and Interferometry, is, therefore, strongly dependent on the ability to select the scientific programs to be run during a night and the set-up of instrumentation to be used during each night. This selection and management depends on the atmospheric conditions and in particular on the optical turbulence conditions ($\CN2$ profiles) and is called, in the astronomical context, 'flexible-scheduling'. All the top-class telescopes and future generation telescopes (Extremely Large Telescopes - ELTs) are planning to use the Service Mode to schedule the scientific programs. Such a mode takes into account the status of the atmospheric conditions beside to the quality of the scientific programs and this permits to concretely perform the flexible scheduling. As extensively explained precedently \citep{masciadri2013}, the Service Mode is crucial and mandatory for an efficient exploitation of the best ground-based astronomical facilities. 

The idea to reconstruct $\CN2$ profiles, with mesoscale non-hydrostatical models has been originally proposed by \citet{masciadri1999}. The authors proposed a parameterization of the optical turbulence employing the prognostic equation of the turbulent kinetic energy (TKE). These models are certainly the most suitable models to be used for this kind of applications mainly because the General Circulation Models (GCMs), that are extended on the whole globe, have necessarily a lower horizontal resolution as extensively explained in \citet{masciadri2013}. This approach has been followed by successive developments in many other studies using the Astro-Meso-Nh code (\citet{masciadri2001}; \citet{masciadri2002}; \citet{masciadri2004}; \citet{masciadri2006}; \citet{lascaux2010}; \citet{hagelin2011}; \citet{lascaux2011}; \citet{masciadri2017}) that, over the years, contributed to prove that $\CN2$ and integrated astroclimatic parameters can be reliably forecasted for astronomical applications (at mid-latitudes as well as at polar latitudes) following this approach. 

In most recent years other studies concerning the OT forecast on the whole atmosphere have been published using other mesoscale models (\citet{trinquet2007}; \citet{cherubini2008};\citet{cherubini2011}; \citet{giordano2013}; \citet{liu2015}) or using GCMs (\citet{ye2011}; \citet{osborn2018}). Methods employed include the TKE prognostic equation approach and empirical approaches based on description of the $\CN2$ as a function of the temperature and wind speed. Studies with GCMs are all performed with empirical approaches.
A very recent analysis \citep{masciadri2019} confirmed the thesis that mesoscale models provides better performances than GCMs in the estimate of the seeing. A different study \citep{turchi2019} showed that a gain is obtained with mesoscale models with respect to GCMs by forecasting the precipitable water vapor. We remind that mesoscale models have been invented exactly to bypass intrinsic limitations of the GCMs. This is not therefore surprising. 

The most recent version of the Astro-Meso-Nh model has been used to set-up an automatic and operational forecast system for the OT and some relevant atmospheric parameters with the goal to support the observations of the Large Binocular Telescope (LBT) located at Mt.Graham (US) (\citet{masciadri1999}; \citet{masciadri2017}). LBT is a binocular telescopes, with two 8.4 m primary mirrors working in interferometric configuration; it is therefore considered the precursor of the Extremely Large Telescopes. The operational forecast system is called ALTA Center\footnote{\href{http://alta.arcetri.inaf.it}{http://alta.arcetri.inaf.it}. Also accessible through lbto.org.} (Advanced LBT Turbulence and Atmosphere Center), it is running since a couple of years and it is in continuum evolution.  The Mauna Kea Weather Center{\footnote{\href{http://mkwc.ifa.hawaii.edu/}{http://mkwc.ifa.hawaii.edu/}} is, at our knowledge, the only other similar tool existing at present time. 

The approach that our team followed so far, for an operational application (see ALTA Center), consists on calculating the forecast of the OT for the next night taking care to provide the forecast a few hours before the beginning of the night, typically in the early afternoon. Hereafter, we will call this as 'standard' strategy or 'standard' configuration. Results obtained so far with this approach are very promising \citep{masciadri2017}. The technique we proposed and implemented in ALTA Center has a few important appealing characteristics: \\
(1) the accuracy of the forecast system for the OT (or equivalently the RMSE) is of the same order of the accuracy attainable with instruments. In other words, the dispersion between prediction and observations is comparable to the dispersion of observations obtained with different instruments (for example \citet{masciadri2017}); \\
(2) it permits to have a temporal frequency of the forecast of 2 minutes (but this can be further reduced in case of necessity). This feature makes mesoscale models more attractive with respect to GCMs having a frequency from 1 hour to 6 hours; \\
(3) it permits to implement operational forecast systems with mesoscale models without the necessity of expensive clusters i.e. with a relative cheap approach preserving the best model performances. 

In this paper we started from the consideration that, if we take into account the overhead necessary to carry out a scientific program and/or the logistic to switch the beam from an instrument to another one on top-class telescopes, the most critical time scale on which to optimise observations supported by AO is of one or two hours.  It should be, therefore, very useful to have forecasts on this time scale and to know if we can improve model performances with respect to the standard strategy (characterised by longer time scales).  
The question is therefore: is it possible to achieve this goal using filtering techniques such as autoregression, Kalman filter or neural networks (also known as machine learning techniques) ? The idea behind this is that the knowledge of in-situ measurement might help in eliminating some short time scales biases and trends that affect the forecast of atmospheric models at longer time scales. In a preliminary analysis \citep{turchi2018} our team showed that such an approach might be promising. 

In this paper we concentrated our attention on the auto-regressive technique that depends simultaneously on a continuous data-stream of real-time measurements taken in-situ and on time series of the atmospherical model outputs. We decided to start with the auto-regressive method because the astronomical application implies the interest for a specific point, the location of the telescope. It is highly possible therefore that observations done in just one location can be enough to achieve our objective. We considered here the Astro-Meso-Nh forecasts done using the standard strategy available in the early afternoon. We defined the algorithm for the auto-regressive forecast, we carried out a complete quantitative analysis on the impact of such technique on the forecasts of the seeing and other relevant atmospheric parameters and we defined the best configuration to obtain the highest gain i.e. the best model performances. We finally implemented this system in the automatic and operational forecast system ALTA Center that has therefore, now, the possibility to provide forecasts at different time scales: the forecasts of the next night on a time scale of the order of six to fifteen hours and a forecast at short time scale i.e. order of one hour.

The plan of the paper is synthesised here. In Section \ref{obs} we described the observations and in Section \ref{model} the configuration of the atmospherical model used for this study. In Section \ref{AR} it is reported the principle of the auto-regression method proposed and analysed in this paper. Section \ref{results} reports the results of the auto-regressive technique (AR) performances in forecasting the various parameter using statistical operators of different nature applied to a statistical sample of one year. In order to quantify the impact of the AR method on the forecasting performances, we compare these results with respect of the 'method by persistence', i.e. the simple use of real-time measurements. In Section \ref{auto} we describe the implementation of the method in the operational forecast system ALTA Center and finally, in Section \ref{concl}, the conclusions and perspectives are reported.

\begin{figure}
\begin{center}
\includegraphics[]{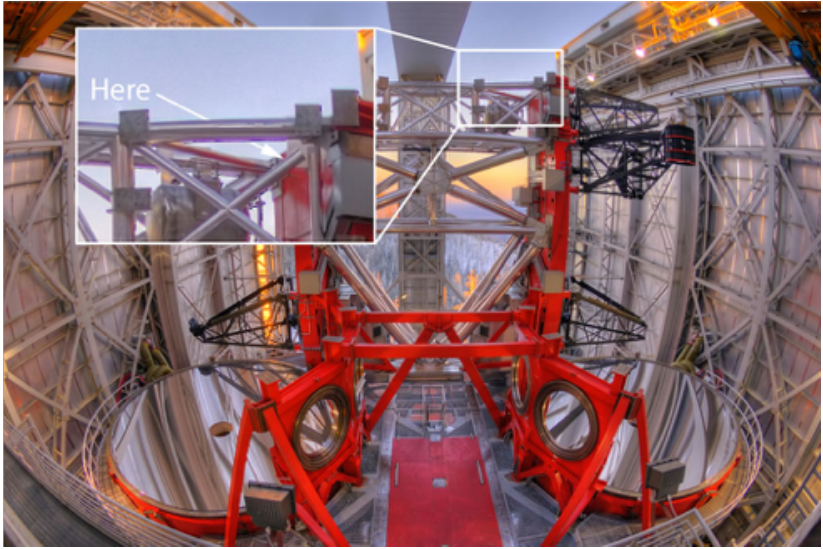}\\
\end{center}
\caption{\label{fig:dimm} Location of the DIMM running nightly at Mt.Graham and measuring the seeing. The instrument is located on the top of the LBT dome (see zoom in the square with white frame.)}
\end{figure}


\section{Observations}
\label{obs} 

Different typologies of observations have been considered as a reference. For the atmospheric parameters we considered the real-time measurements routinely done with sensors placed on the roof of the telescope dome and successively stored in the LBT telemetry.  As described in \citet{turchi2017}, the sensors are installed on masts having different heights and located on the LBT roof (53~m above the ground). Temperature (T) and relative humidity (RH) sensors are located at 55.5~m above the ground (sensor at 2.5~m above the roof). Wind speed (WS) and wind direction (WD) are both measured by two different anemometers placed in two different locations on the roof that we call 'front' and 'rear' (at 56~m and 58~m above the ground and 3 and 5 meters above the roof). WS measurements are computed using a combination of measurements taken by the two sensors using an algorithm that takes into account the relative position of the telescope line-of-sight with respect to the wind direction. We refer the reader to \citet{turchi2017} for a detailed description of the algorithm. We considered only WD measurements taken from the rear anemometer because we verified that rear and front WD measurements are statistically equivalent. Observations are stored with a frequency of around one second in the LBT telemetry. For the seeing i.e. the integral of the $\CN2$ on the whole atmosphere, we considered the measurements taken with a Differential Image Motion Monitor (DIMM) i.e. a monitor installed inside the LBT dome, close to the roof (Fig.\ref{fig:dimm}) that nightly monitors the turbulence affecting the quality of images on the scientific camera. Looking at the position of the DIMM inside the telescope dome, we deduce that this instrument necessarily measures also the dome seeing (if any). On the other side, this represents the real turbulence affecting the images obtained at the focus of the telescope.  Even if the LBT-DIMM does not measure the 'pure' atmospheric turbulence, it provides a more realistic estimate of the turbulence affecting the images. It has been decided therefore to assume the DIMM measurements as our reference and use these estimates for the model validation. The elimination of the dome seeing should not be trivial considering the information that are accessible but it is visibly not really very relevant. Assuming the DIMM as a reference means that we are calibrating the model to take into account a surplus of turbulence due to the dome so that the predicted turbulence is equivalent to the total turbulence affecting in reality the camera. This is obviously done in statistical terms. It is important to note that, at present, at Mt.Graham there isn't a vertical profiler running nightly. This means that there are no real-time measurements of the wavefront coherence time ($\tauO$) and the isoplanatic angle ($\thetaO$). Both parameters depend, indeed, on the integral of the $\CN2$ on the atmosphere. We will treat therefore in this study only the seeing as integrated astroclimatic parameter. 

\begin{table}
\caption{Astro-Meso-NH model grid-nesting configuration. In the second column the horizontal resolution $\Delta$X, in the third column the number of horizontal grid-points, in the fourth column the domain extension.}
\begin{center}
\begin{adjustbox}{max width=\columnwidth}
\begin{tabular}{|c|c|c|c|}
\hline
Domain & Grid   & Domain size & $\Delta$X  \\
       & Points & (km)        & (km)  \\
\hline
Domain 1 &  80$\times$80  & 800$\times$800 & $\Delta$X = 10\\
Domain 2 &  64$\times$64  & 160$\times$160 & $\Delta$X = 2.5\\
Domain 3 & 120$\times$120 &  75$\times$50  & $\Delta$X = 0.5\\
Domain 4 & 100$\times$100 & 10$\times$10 & $\Delta$X = 0.1 \\
\hline
\end{tabular}
\end{adjustbox}
\label{tab:orog}
\end{center}
\end{table}

\begin{figure*}
\begin{center}
\includegraphics[width=0.9\textwidth]{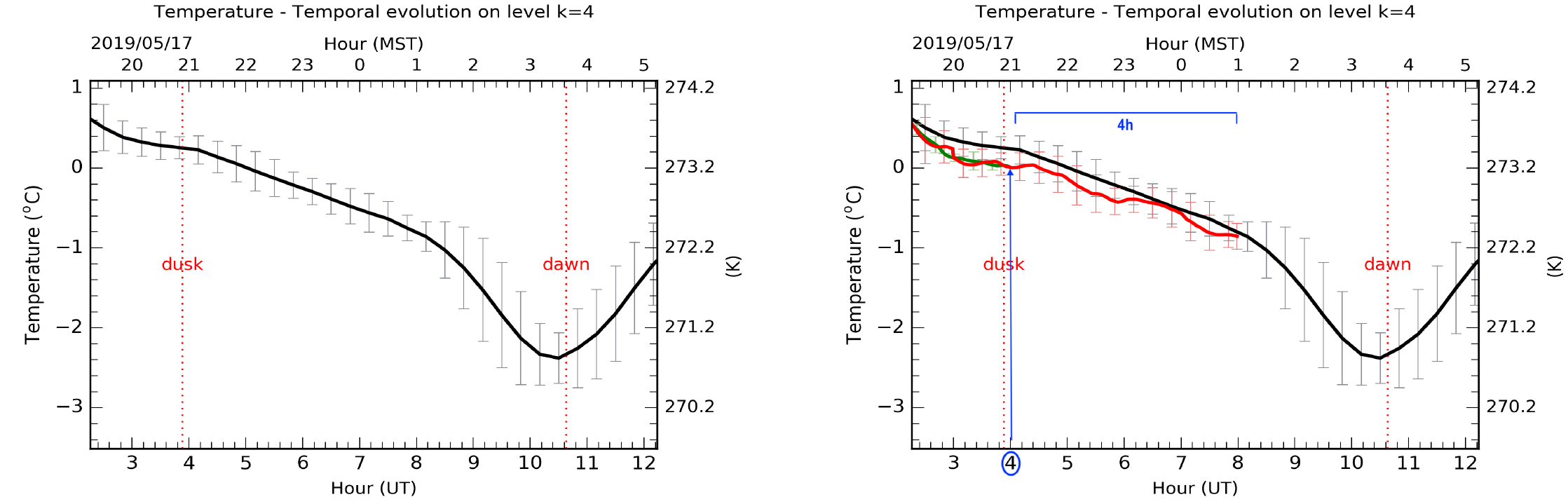}\\
\end{center}
\caption{\label{fig:AR_princ} Left: temporal evolution of the forecast of the temperature for the whole night (2019/05/17) at Mt.Graham. The forecast is available at 14:00 UT of the day before. On the x-axes is reported the time expressed in UT (bottom) and local time (top). Right: temporal evolution of the forecast of the temperature available at 14:00 UT of the day before (black line); real-time measurements in situ (green-line); forecast of the temperature using the AR technique (red line). The latter is calculated at 04:00 UT and extended on the successive four hours (see text).}
\end{figure*}

\begin{figure*}
\begin{center}
\includegraphics[width=1\textwidth]{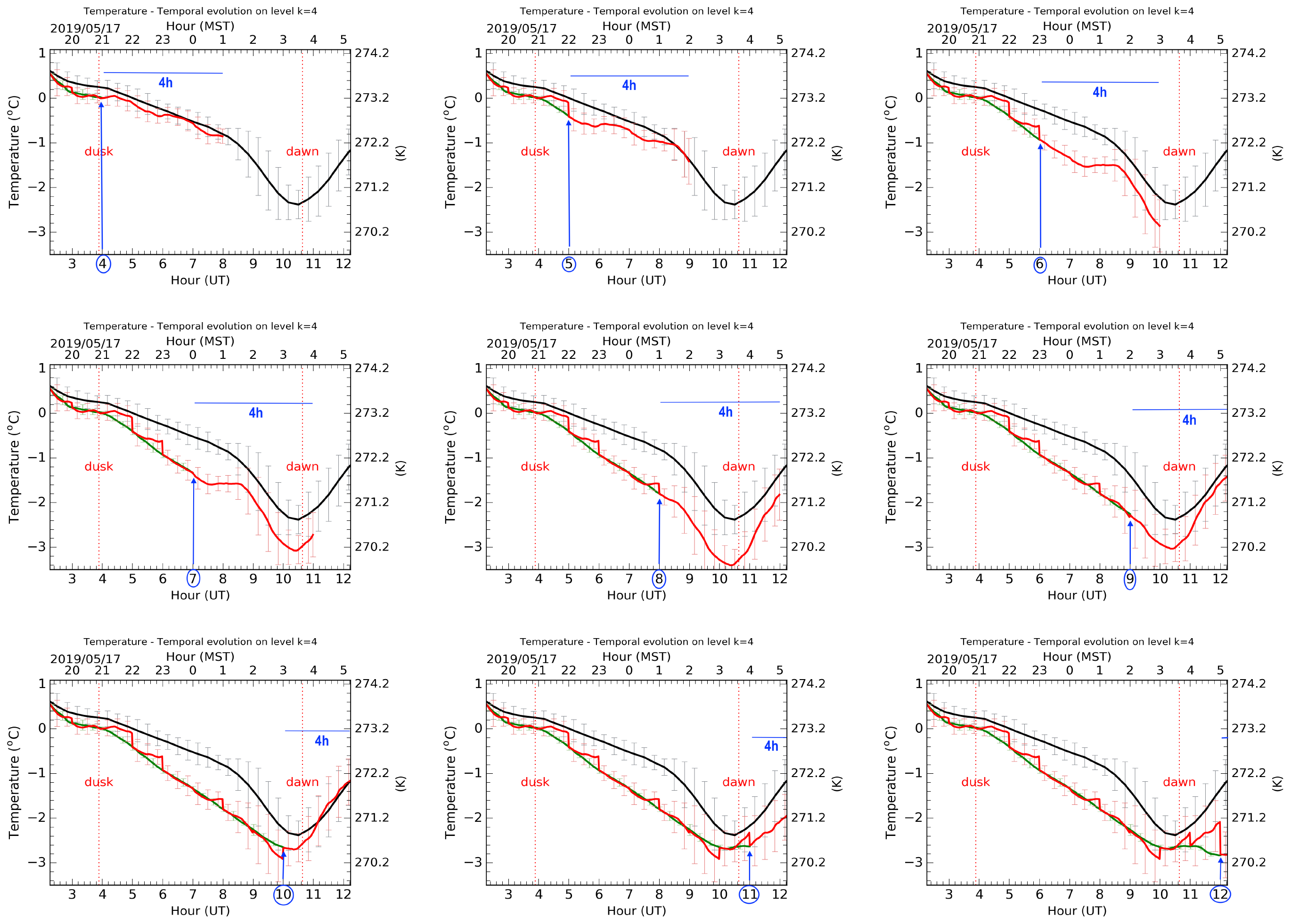}\\
\end{center}
\caption{\label{fig:AR_oper} Temporal sequence of the up-dated forecasts of the temperature with one hour step during the night 2019/05/17. First image on top-left is the situation at 04:00 UT of 2019/05/17, last image on bottom right is the situation at 12:00 UT of 2019/05/17. The sequence is read by rows, from the left to the right. The black line is the forecast of the temperature available at 14:00 LT of the day before. It is therefore always the same in all the pictures. The green-line is the real-time measurements. In each picture the end of the green line ends at the time in which the AR forecast is calculated. The red-line is the forecast of the temperature obtained with the AR technique. The red line in the last picture (bottom-right) represents the model forecast at 1h for the whole night.}
\end{figure*}


\section{Model}
\label{model}

The atmospherical mesoscale model Meso-Nh\footnote{\href{http://mesonh.aero.obs-mip.fr/mesonh52}{http://mesonh.aero.obs-mip.fr/mesonh52} - we used the Masdev5.2 version of the code.} (\citet{lafore1998}; \citet{lac2018}) has been used in this study  for the forecast of the atmospheric parameters (T, RH, WS and WD) while the Astro-Meso-Nh code (\citet{masciadri1999}; \citet{masciadri2017}) has been used for the forecast of the optical turbulence i.e. the seeing. In both cases it is possible to retrieve the spatio-temporal evolution of three, two or mono dimensional parameters over a specific limited area of the Earth. In the case of the seeing, the model calculates first a 3D map of the $\CN2$ in a region around the telescope, and afterwards, the $\CN2$ is integrated on the whole atmosphere ($\sim$ 20km a.g.l.) to obtain the seeing i.e. a 2D map. The same model configuration described in \citet{turchi2017} has been implemented. We synthesise here the main elements to permit the readers to follow. For what concerns the Meso-Nh model, the system of hydrodynamic equations is based upon an anelastic formulation that permits an effective filtering of acoustic waves. The model uses the \citet{gal1975} coordinates system on the vertical and the C-grid in the formulation of \citet{arakawa1976} for the spatial digitalization. In this study, we used in the wind advection scheme the 'forward-in-time' (FIT) numerical integrator instead of the' leap-frog' one. Such a solution allows for longer time steps and therefore shorter computing time. The model employs a one-dimensional 1.5 turbulence closure scheme \citep{cuxart2000} and we used a one-dimensional mixing length proposed by \citet{bougeault1989}. The surface exchanges are computed using the interaction soil biosphere atmosphere (ISBA) module \citep{noilhan1989}. The seeing ($\varepsilon$) is calculated with the Astro-Meso-Nh code developed by \citet{masciadri1999} and since there in continuous development by our group. The geographic coordinates of Mt.Graham are (32.70131, -109.88906) and the height of the summit is 3221~m above the sea level. We used a grid-nesting technique \citep{stein2000} consisting in using different embedded domains of the digital elevation models (DEM i.e. orography) extended on smaller and smaller surfaces, with progressively higher horizontal resolution but with the same vertical grid. Simulations of the OT are performed on three embedded domains centered on the summit where the horizontal resolution of the innermost domain is $\Delta$X = 500~m (Table \ref{tab:orog}). We used four domains and a highest resolution of 100~m (Table \ref{tab:orog}) for the WS because such a configuration better reconstructs the WS close to the surface when the WS is strong. The model is initialised with analyses provided by the General Circulation Model (GCM) HRES of the European Center for Medium Weather Forecast (ECMWF) having an intrinsic horizontal resolution of around 9 km. All simulations we performed with Astro-Meso-Nh start at 00:00 UT of the day J and we simulate in total 15 hours\footnote{To avoid misunderstandings found in the literature, we highlight that a simulation of 15h does not mean that we need 15h to simulate that period. It means that we reconstruct the atmospheric evolution of 15h. The simulated time and the effective calculation time required to perform a calculation are two different concepts of 'time'.}. We consider data starting from the sunset up to the sunrise. During the 15 hours the model is forced each six hours (synoptic hours) with the forecasts provided by the GCM related to the correspondent hours. We consider the $\CN2$ outputs with a temporal frequency of two minutes. All the other atmospheric parameters have a temporal frequency of the order of the second. In all cases the simulated data are extracted from the innermost domain (domain 3 or 4 - see discussion on the wind speed a few line above). We have a 54 vertical levels with a first grid point of 20~m, a logarithmic stretching of 20 per cent up to 3.5~km above the ground and almost constant vertical grid size of $\sim$600~m up to 23.57~km.  The height of the first grid point has been fixed to be able to resolve the in-situ measurements of the various parameters analysed in this study. 


\section{Autoregressive method}
\label{AR}

As we said in Section \ref{intro}, the goal of this study is to verify if we can improve the model performances of forecasts on time scales of a few hours. The method that we propose to use in this paper is based on the auto-regressive (AR) technique. We chose a formulation inspired by \citet{dzhaparidze1994}.  The method is based on a function that depends on the difference between the real-time observations taken in-situ that we take as a reference (i.e. we assume to be the "truth") and on the forecasts performed by the atmospheric model. When we deal about 'atmospherical model' we are referring to the forecast of the model in standard configuration (see Section \ref{model}) that is available early in the afternoon of the day before\footnote{For simplicity, we will call hereafter simply 'model' the Meso-Nh or the Astro-Meso-Nh models, knowing that the first one is used for the atmospheric parameters, the second one for the OT.}. The auto-regressive model (AR) $X_{t+1}^{*}$ calculated at the (t + 1) is:

\begin{equation}
X_{t+1}^{*}=M_{t+1} + X_{t+1}
\label{eq:ar}
\end{equation}

\noindent
where M is the model output at the time (t +1) and the function X at the time (t + 1) depends on the difference between the observations and the atmospherical model outputs calculated on a polinomial function built with the addition of P terms characterised by P coefficient $a_{i}$ (called regressors) in the form:

\begin{equation}
X_{t+1}=\sum_{i=1}^{P}a_{i}(OBS_{t-i+1}-MOD_{t-i+1}).
\label{eq:ar1}
\end{equation}
\noindent

where the variable $OBS$ indicates the real-time measurements and $MOD$ the atmospheric model outputs in the standard configuration. From one side, the larger is P, the larger is the number of the regressors, the more accurate is the fit to the trend of the past observations. On the other side, we have interest in limiting the number of the coefficients $a_{i}$ to limit the computation time. We identified an optimal trade-off P=50 for a temporal frequency of 1 minute.

The values of the 50 regressors is obtained through a Least Mean Square (LSM) method applied to a finite number of nights in the past i.e., for example, the last 3, 4, 5, etc. nights.

Figure \ref{fig:AR_princ} shows how the AR method works. The picture shows, as an example, the forecast of the temperature but the same procedure can be used for whatever parameter. On the left side is shown the standard forecast of the night 2019/05/17\footnote{The date refers to the start of the night.} that is available early in the afternoon. On the right side is reported an example of the AR method applied at 04:00 UT. The black line is the standard forecast with the atmospheric model (same as the left side), the green line represents the real-time measurements up to 04:00 UT (that is the present time), the red line represent the forecast calculated at 04:00 UT with the AR method for the successive 4 hours. As we are interested here on studying the forecast performances on time scales of one or two hours, we considered therefore a AR forecast of 4 hours that certainly covers this time scale. We expect that the effect of the data-assimilation of the local measurements provides an improvement of the forecast that is maximum close to the present time (nowcasting) and it decreases with the time up to disappear. The positive effect of the AR method vanishes after a $\Delta$T that is when the performance of the AR method is equal to the performance of the atmospheric model in standard configuration. Later on, in Section \ref{results}, this aspect will be treated in a more detailed way.  If the same procedure described in Fig. \ref{fig:AR_princ} is repeated with the suitable frequency during the whole night, it is possible to obtain a forecast on a time scale of one hour. 

Figure \ref{fig:AR_oper} reports the sequence of successive AR forecasts that are recalculated at each full hour during the night. The sequence has to be read from the top to the bottom, from the left to the right, following the different rows. We observe that, in each successive picture of the sequence, the green line becomes longer of one hour and the red-line, showing the forecast related to the successive four hours,  shifts of one hour on the right. If we consider the red line of the last picture (bottom-right) extended on the whole night, we have the performance of the system on a time scale of one hour. We highlight that, in this computation and procedure, we take into account only data between the sunset and the sunrise. 

As said previously, the unique free parameter remains the number of nights (N) on the past on which to calculate the values of the regressors. As we will see later on, N=5 is a suitable number for our application. The whole analysis presented in Section \ref{results} has been performed assuming this value for N.  

\begin{figure*}
\begin{center}
\includegraphics[width=0.9\textwidth]{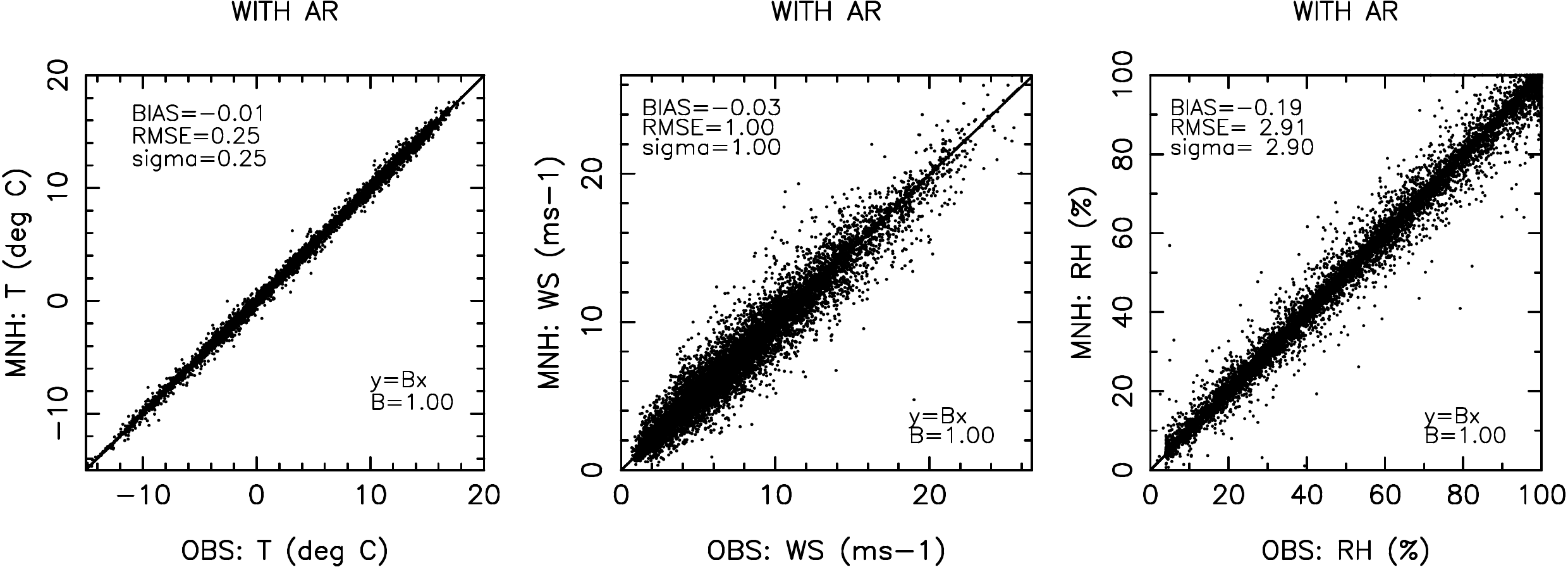}\\
\end{center}
\caption{\label{fig:stat} Scattering plot between observations and AR method outputs for absolute temperature (left), wind speed (centre) and relative humidity (right). Data are treated with a moving average on one hour and resampling on 20 minutes. Number of nights on which the regressors are calculated is N=5.}
\end{figure*}

\begin{figure*}
\begin{center}
\includegraphics[width=0.9\textwidth]{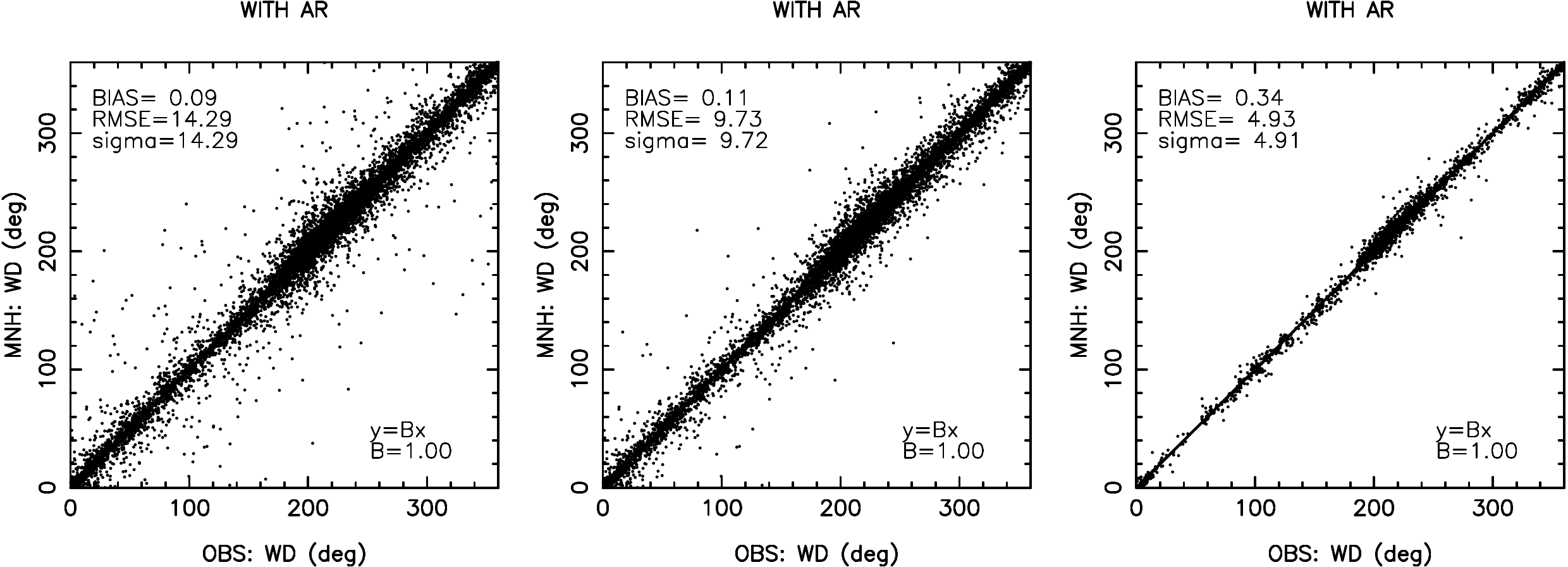}\\
\end{center}
\caption{\label{fig:stat_WD} Scattering plot between observations and AR method outputs for the wind direction (left). Same results but filtering out all cases in which WS is weaker than 3 ms$^{-1}$ (centre) and weaker than 10 ms$^{-1}$ (right). N=5 as in Fig. \ref{fig:stat}.}
\end{figure*}

\begin{figure*}
\begin{center}
\includegraphics[width=0.9\textwidth]{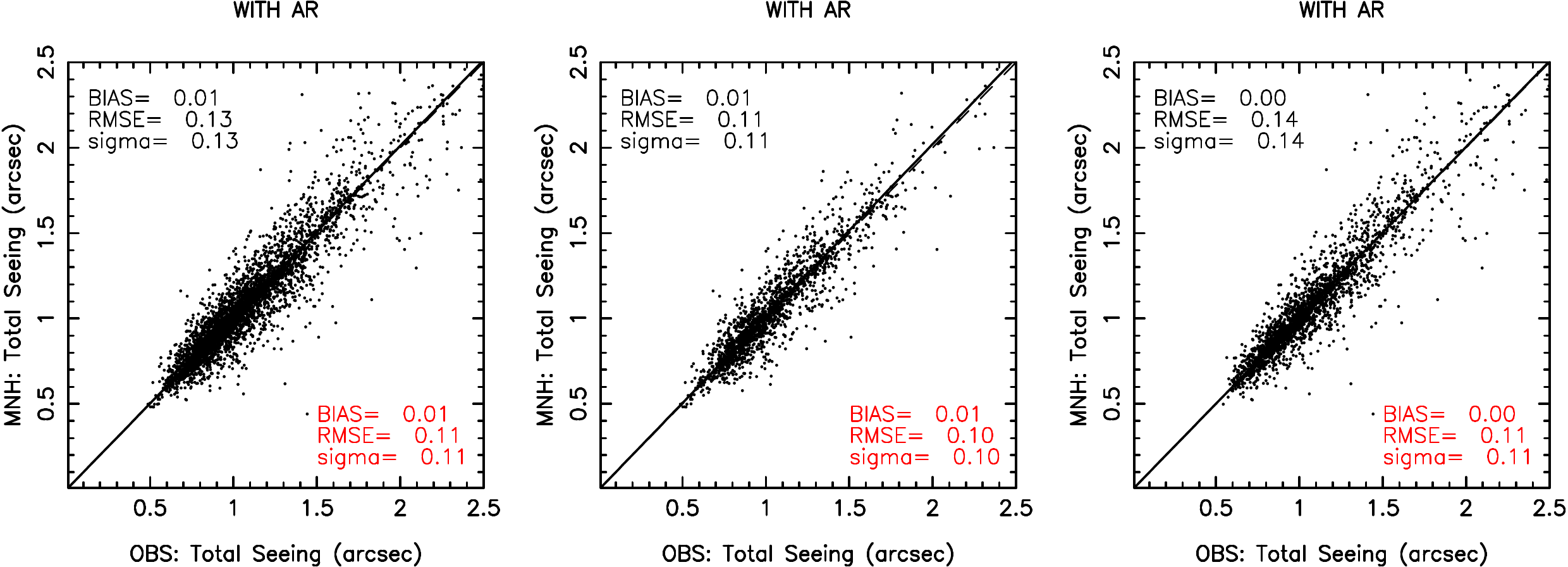}\\
\end{center}
\caption{\label{fig:stat_see} Scattering plot between observations and AR method outputs for the total seeing calculated on the whole year (left), in the summer period (centre) and in the winter period (right). Summer period is included in the [April-September] interval, winter period in the [October-March] interval. In black results considering all values, in red considering only observations below 1.5" (see discussion in the text). N=5 as in Fig. \ref{fig:stat}.}
\end{figure*}


\section{Results}
\label{results}

In order to quantify the model performances of the forecasts on one hour time scale using the AR method built as described in Section \ref{AR} it is necessary to consider a very rich statistical sample because the AR method requires a sequence of observed data related to successive nights in which it is important to minimise the number of breaks (lack of measurements). We considered therefore data of all the nights of the whole year 2018 and we calculated the statistical operators (bias, RMSE and $\sigma$)\footnote{We refer the reader to \cite{masciadri2017} for the definition of the statistical operators.} for temperature, relative humidity, wind speed, wind direction and the total seeing. Real-time measurements and outputs of the atmospheric model in standard-configuration related to these parameters have been treated using the same procedure: we first apply a moving average of one hour to filter out the high frequencies and put in evidence the forecast trend, we perform a resampling on a time scale of 20 minutes\footnote{A resampling on 10 minutes provide a very similar result.} and we conclude with the calculation of the various statistical operators. 

Figure \ref{fig:stat} shows the scattering plot related to the temperature (left), the wind speed (centre) and the relative humidity (right) obtained with an AR at a time scale of 1 hour. Figure \ref{fig:stat_WD} shows the scattering plot of the WD at the same time scale of 1 hour obtained including all the data (left), filtering out all the data associated to wind speed weaker than 3 ms$^{-1}$ (centre) and filtering out all data having a wind speed weaker than 10 ms$^{-1}$. We skipped out the data associated to a WS weaker than 3 ms$^{-1}$ because under this condition it is extremely difficult (and meaningless) to quantify the WD because of the high variability of the WD. The central picture of Fig.\ref{fig:stat_WD} is therefore more representative for the WD than the left one. We skipped-out data weaker than 10 ms$^{-1}$ to quantify the model performances in those cases that are certainly the most critical one for the ground-based observations, i.e. those in which the WS is very strong. To conclude, Fig. \ref{fig:stat_see} shows the scattering plot for the seeing in the whole year (left), in the summer [April - September] interval (centre) and winter [October - March] interval (right). 

%
%

\begin{table*}
\begin{center}
\caption{\label{tab:perf_rmse} RMSE as obtained with the atmospheric model in the standard configuration and as obtained with the AR method on a 1h time scale. In the case of the seeing we considered only seeing below 1.5". This threshold is more than representative for the AO applications and it guarantees a model performances comparable to the dispersion obtained with measurements.}
\begin{tabular}{lccccc}
\hline
   RMSE   & T  & RH  & WS & WD ($>$ 3ms$^{-1}$) & Seeing \\
                 & ($^{\circ}$ K) & ($\%$) & (ms$^{-1}$) & (degrees)   &  (arcsec) \\
\hline
 atm. model standard config. &  0.98& 14.17 & 2.81 & 34.71 &  0.30 \\
\hline
 AR (@ 1h) & 0.25  &  2.91 & 1.00 &  9.73 & 0.11  \\
 \hline
\end{tabular}
\end{center}
\end{table*}

\begin{table}
\begin{center}
\caption{\label{tab:gain} First row: gain obtained for the RMSE for the different atmospheric and astroclimatic parameters of the AR method on a time scale of 1h with respect to the model standard configuration. Second row: gain using the method by persistence on the same time scale with respect to the model standard configuration. Third row: gain of the AR method with respect to the method by persistence on the same time scale.}
\begin{tabular}{lccccc}
\hline
  GAIN    & T  & RH  & WS & WD & Seeing \\
\hline
AR   &  3.90 & 4.90 & 2.80  & 3.60  & 2.70  \\
 Persistence  &  2.40 & 3.00 & 1.80  & 2.40 &   2.00   \\
 AR / Persistence &  1.63 & 1.63 & 1.56 & 1.50 &  1.35   \\    
\hline
\end{tabular}
\end{center}
\end{table}

Table \ref{tab:perf_rmse} reports the RMSE obtained for the AR method at a time scale of 1 hour and with the atmospheric model in the standard configuration. As we observe that, in the standard configuration\footnote{This feature has not been observed after the application of the AR technique as one can see in Fig.\ref{fig:stat_see}}, the dispersion of the seeing increases for large seeing values but the forecasts are less interesting for those cases. We decided, therefore, to consider observations below 1.5". From a practical point of view, indeed, in the astronomical context it is poorly interesting to discriminate seeing values between 1.5" and larger values. We maintained both cases for the AR (Fig.\ref{fig:stat_see}) because the RMSE are very similar. We observe that, for all the parameters, the values of RMSE obtained with the AR method at a time scale of 1h are definitely better than for the standard configuration with consistent gains that are variable depending on the parameters between a minimum of a factor 2.7 and a maximum of 4.9 (Table \ref{tab:gain}-first row). Those gains are definitely consistent and, at our knowledge, these model performances have never been achieved before. 
In Annex \ref{annex_a} is reported a detailed description on the number of nights used to analyse this statistics for each parameter. 
The extremely small value of the RMSE for the temperature of the order of 0.25$^{\circ}$ tells us that, with such performances in predicting the temperature close to the ground, the elimination of the dome seeing through a thermalisation of the primary mirror temperature and the atmosphere inside the dome with respect to the external temperature is not a dream anymore, as declared by \citet{racine1991}. 

It remains to consider how to fix the number of nights on which to calculate the regressors. Figure \ref{fig:gain} shows how the RMSE obtained with the AR method changes as a function of the interval of time $\Delta$T on which we calculate the forecast and as a function of N. We decided to consider $\Delta$T = 1 hour as a minimum value because, considering the logistic requiring a change of program or the set-up of an instrument, makes poorly  interesting to go below this threshold. As expected, the gain is maximum at 1h and it decreases as $\Delta$T increases\footnote{When $\Delta$T = 0 we have the nowcasting.}. The black line, represents, for each parameters, the RMSE obtained with atmospherical model in standard configuration that is obviously constant for the whole night. We note that there is a saturation effect for N equal to 4 or 5. We decided therefore to use N=5 in our calculation because no further gain is visible for N larger than 5. The point in which coloured lines cross the black line represents the $\Delta$T at which the AR stops to present an improvement in the performances with respect to the standard configuration and it starts to diverge. For $\Delta$T larger than this threshold, the standard configuration is more advantageous than the AR method. This is exactly the expected trend as the in-situ measurements stop to have a positive influence on the forecast performances for $\Delta$T too large. We can observe that, the AR continues to maintain a gain different from zero up to a time scale of the order of 4-6 hours\footnote{The reason why we display the figure only up to 4 hours is to avoid a too large inhomogeneity in the statistical representativity of the samples. The number of samples for each $\Delta$T decreases indeed, as we increase $\Delta$T.}. \\

\begin{figure*}
\begin{center}
\includegraphics[width=0.9\textwidth]{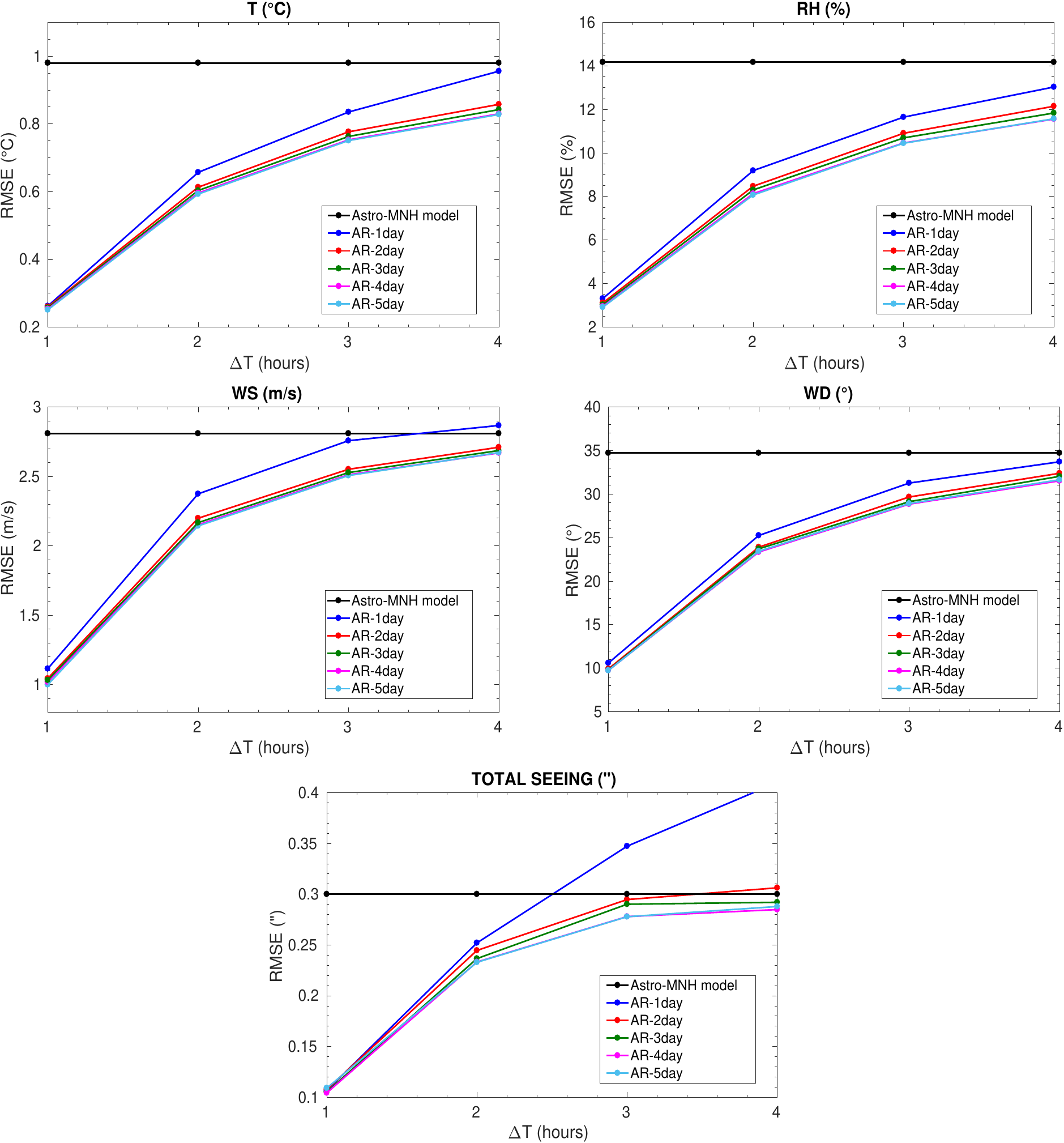}\\
\end{center}
\caption{\label{fig:gain} Dependency of the RMSE of different atmospheric parameters with respect to different $\Delta$T. On the x-axes the 'forecast time' $\Delta$T = (T$_{f}$ - T$_{i}$) where T$_{i}$ is the time in which the forecast is calculated and T$_{f}$ is the time for which the forecast refers to. Example: $\Delta$T = 1 means a forecast at 1 hour calculated at T$_{i}$. Top-left: temperature; Top-right: relative humidity; Centre-left: wind speed; Centre-right: wind direction of all data for which WS > 3ms$^{-1}$; Bottom: total seeing (we considered observations below 1.5"). The horizontal black line represents the RMSE calculated with the model in standard configuration. }
\end{figure*}

\begin{table}
\begin{center}
\caption{\label{tab:clim} Climatology tertiles calculated on measurements extended on one full solar year (2018) for the absolute temperature T, the wind speed WS, the relative humidity RH and the seeing.}
\begin{tabular}{lccc}
\hline
  Param.    & 1st tert.  & Median  & 3rd ter.  \\
\hline
T ($^{\circ}$) & 1.00 & 4.11  & 8.23 \\
WS (ms$^{-1}$) & 5.52 & 7.15  & 9.08 \\
RH ($\%$) & 31.63 & 47.48 & 66.67 \\
seeing (") & 0.93 & 1.05 & 1.20  \\
seeing (< 1.5") & 0.90 & 0.99 & 1.10  \\
\hline
\end{tabular}
\end{center}
\end{table}

Once analysed the gain obtained employing the auto-regression approach, it might be interesting to quantify which is the gain on a time scale of a few hours if we use just real-time measurements instead of the filtering techniques. We call this approach 'method by persistence'. This means that, at each full hour, the forecast extended on the successive 4 hours, is obtained by considering the present time measurements as a constant for all its future evolution. Fig.\ref{fig:ar_vs_pers} shows the RMSE versus the $\Delta$T obtained with the optmized AR method (N=5) and the persistence method. It is clearly visibly that, as expected, even if the use of pure real-time measurements provides an improvement of the forecast performances on short time scales with respect to the standard configuration of the model, the AR method that we propose has definitely a more important gain and better performances for all the atmospheric parameters including the optical turbulence with differences (with respect to the persitence method) that are quantitatively not negligible. Table \ref{tab:gain}-second row reports the gain of the persistence method with respect to the model forecast in standard configuration. Table \ref{tab:gain}-third row reports the gain of the AR method with respect to the persistence method. Looking at Fig.\ref{fig:ar_vs_pers} it is also possible to observe that, in the case of the AR method, the gain persists for a much longer $\Delta$T with respect to the persistence approach. It is worth to note that, of course, the black line of the model forecast in standard configuration is available much earlier than the start of the observing night. It is therefore obviously worse with respect to the other two methods. The fair comparison is therefore between the red and the blue lines.\\

To complete the analysis of the model performances, we finally calculate the contingency tables for each parameter from which we can retrieve the probability of detection (POD), the percentage of correct detection (PC) and the extremely bad detection (EBD). Contingency tables allow for the analysis of the relationship between two or more categorical variables.We refer the readers to \citet{lascaux2015} for a detailed definition and description of this tool. To permit the readers to follow the text we refer to Annex \ref{annex_b} that contains a synthesis of the definitions of the statistical operators. Here we just remind the principal role of the contingency tables. Given a statistical sample of observations and predictions, the contingency tables permit to calculate the number of times in which observations and predictions fall in the same intervals of values. We used 3$\times$3 tables for all the parameters with exception of the WD that requires a 4$\times$4 table as it is a 2$\pi$ periodic parameter. Starting from this distribution it is possible to calculate the probability to detect a specific atmospheric parameter in specific intervals of values, the so called POD$_{i}$, the percentage of correct detection (PC) and the extremely bad detection (EBD). The thresholds of the intervals are calculated from the climatology of in-situ measurements and they are, usually, the first and third tertiles of the cumulative distribution. Table \ref{tab:clim} reports the first and third tertiles calculated on one year (2018) of measurements for the different atmospheric parameters. These values are used as thresholds in this study. Table \ref{tab:cont_temp}, \ref{tab:cont_WS}, \ref{tab:cont_RH}, \ref{tab:cont_WD}, \ref{tab:cont_WD-bis} and Table \ref{tab:cont_see} report the results of POD$_{i}$, PC and EBD for temperature, wind speed, relative humidity, wind direction and seeing in the different configuration: atmospheric model in standard configuration and AR at 1 h time scale. For temperature, WS, RH and seeing, we take i=1,2,3; POD$_{1}$ is the probability to detect values smaller than the first tertile; POD$_{2}$ is the probability to detect values between the first and the third tertiles; POD$_{3}$ is the probability to detect values larger than the third tertile. For the WD we take i=1,2,3,4 and POD$_{1}$, POD$_{2}$, POD$_{3}$ and POD$_{4}$ are, respectively, the probability  to detect a value in the range [0$^{\circ}$, 90$^{\circ}$], [90$^{\circ}$, 180$^{\circ}$], [180$^{\circ}$, 270$^{\circ}$] and [270$^{\circ}$, 360$^{\circ}$]. The same calculation has also been done by rotating the thresholds of 45$^{\circ}$ i.e. 45$^{\circ}$, 135$^{\circ}$ and 225$^{\circ}$. \\
In the case of the seeing, we calculate a contingency table that takes into account an accuracy of 0.2". In reality the dispersion between the seeing measured by different and independent instruments (such as Stereo-SCIDAR\footnote{SCIDAR is for SCIntillation Detection And Ranging} and DIMM) can reach values as high as 0.29" \citet{masciadri2019} but we decided to use 0.2" to be more conservative and because this is a technical specification assumed in some top-class telescopes. 

We observe that, for the AR forecasts at 1h, in the case of temperature, RH, WD and seeing, almost all the POD$_{i}$ are very close to the saturation (values in the [94$\%$, 99$\%$] range) i.e. with small space for further improvements. The PC is also of the same order of magnitude and the EBD basically equal to zero. The wind speed is still very good, only POD$_{2}$ = 83$\%$ but the most important ones (POD$_{1}$ and POD$_{3}$)  i.e. the probability to detect extremely weak and the extremely strong wind speed are $>$ 90$\%$. This tool might therefore be extremely important to face the so called {\it "low wind effect"} i.e. a significant deterioration of image quality observed with high contrast imaging instruments such as SPHERE\footnote{High contrast imaging of the Very Large Telescope located at foci of UT3.} \citep{milli2018} when the wind speed is low or absent. This condition enhances the radiative cooling of the spiders that obstruct the big telescopes pupil, creating air temperature inhomogeneities on the phase across the pupil. For WS $\le$ 4ms$^{-1}$ we calculated that the model is able to reconstruct the WS with an RMSE=0.7ms$^{-1}$. On the other extreme, we calculated that, for WS $\ge$ 10 ms$^{-1}$, the model well reconstructs the WS with a RMSE=1.2 ms$^{-1}$. This means that the method is extremely efficient in predicting the conditions of strong wind speed that represent the main cause of vibration of the adaptive secondaries and/or the primary mirrors.

Looking at the same Tables \ref{tab:cont_temp}, \ref{tab:cont_WS}, \ref{tab:cont_RH}, \ref{tab:cont_WD}, \ref{tab:cont_WD-bis} and Table \ref{tab:cont_see}, we observe that performances of the atmospherical model in standard configuration are weaker than those of the AR at 1h as expected, but still very good. We do not comment further results found in this configuration for the atmospheric parameters as a precedent paper has been dedicated to this aspect \citet{turchi2017}. This calculation has been repeated here (with the same statistical sample used for the AR method at 1h) to be able to quantify the improvement in terms of model performances on short time scales. The new result of this paper is however the estimate of statistical operators (POD$_{i}$, PC and EBD) for the seeing (Table \ref{tab:cont_see}-second column) that reveals to be very promising i.e. all the POD$_{i}$ are of the order of 97-98$\%$. 

We put the accent on the most relevant result obtained in this analysis and related to the seeing. The most critical POD$_{1}$ i.e. the probability to detect a seeing weaker than the first tertile is equal to 81$\%$ for the standard configuration and it is equal to 98$\%$ for the AR method at 1h time step. Both are well above the threshold of 33$\%$ that is the percentage that corresponds to the random case and the AR method is very close to the saturation in terms of performances. Somehow weaker is the probability to detect the seeing larger than the third tertile (65$\%$) in the standard configuration as the larger is the seeing, the larger is the dispersion between observations and numerical calculation. We have here more space for further improvements of the technique. 

\begin{figure*}
\begin{center}
\includegraphics[width=0.9\textwidth]{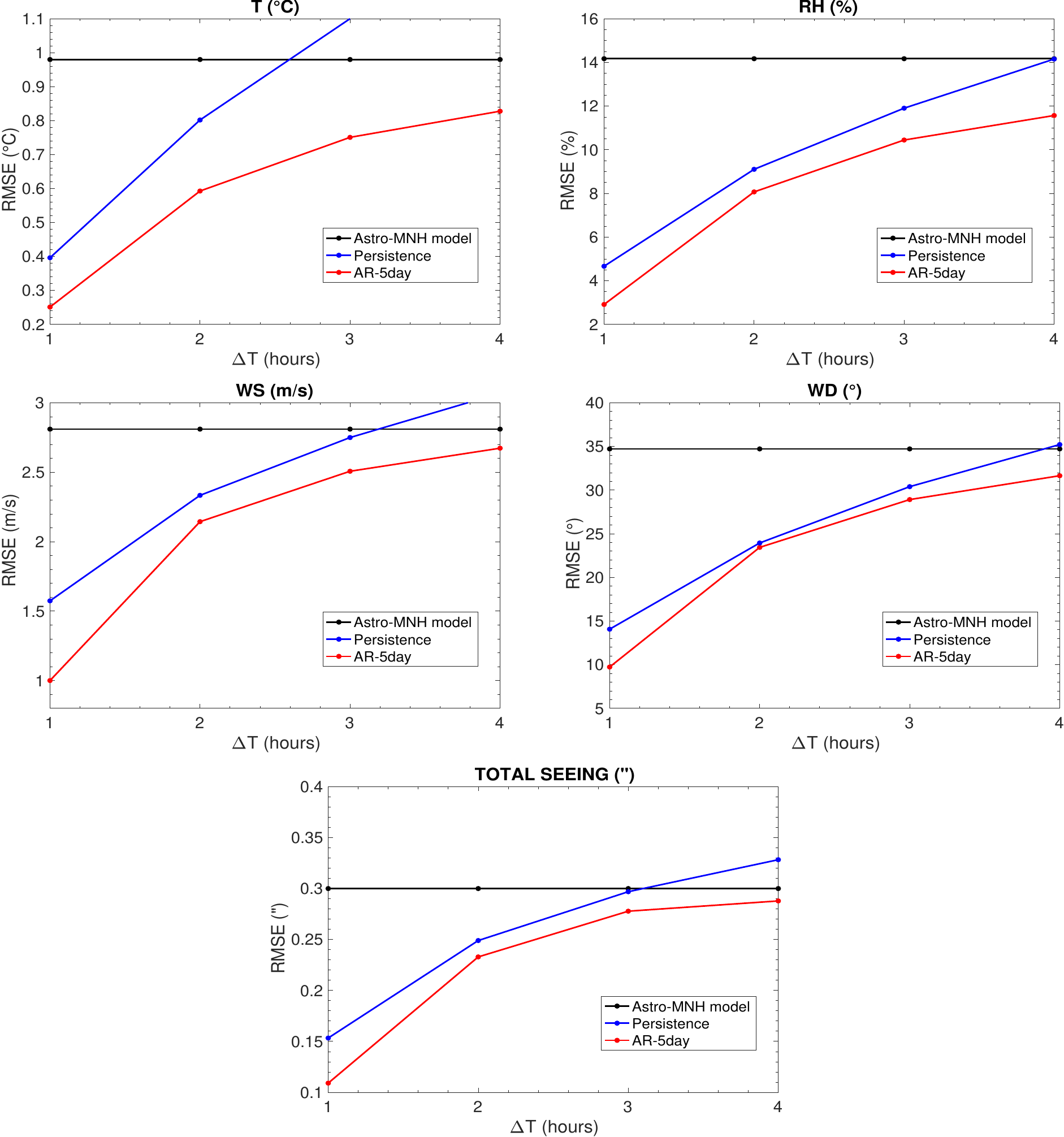}\\
\end{center}
\caption{\label{fig:ar_vs_pers} As Fig.\ref{fig:gain} but are shown the RMSE for the AR method (red line) and the method per persitence (blue line). The horizontal black line represents the RMSE calculated with the model in standard configuration. }
\end{figure*}


\begin{table}
\begin{center}
\caption{\label{tab:cont_temp} Model performances in reconstructing the absolute temperature at different time scales: at 14h i.e. when we provide a forecast early in the afternoon of the day (J-1) for the next night, and at 1h with AR. POD$_{1}$, POD$_{2}$ and POD$_{3}$ are the probability of detection related to the intervals: T $<$ 1$^{st}$ tertile, 1$^{st}$ tertile $<$ T $<$ 3$^{rd}$ tertile, T $>$ 3$^{rd}$ tertile. The 1$^{st}$ and tertiles 3$^{rd}$ are shown in Table \ref{tab:clim}.}
\begin{tabular}{ccc}
\multicolumn{3}{c}{Temperature (T)}\\
\hline
Param. & Forecast   & Forecast with AR  \\
   & the day before ($\%$) & @ 1h ($\%$)\\
\hline
POD$_{1}$ & 96 & 99 \\
POD$_{2}$ & 91 & 98 \\
POD$_{3}$ & 96 & 99 \\
PC & 94 & 99 \\
EBD & 0 & 0 \\
\hline
\end{tabular}
\end{center}
\end{table}

\begin{table}
\begin{center}
\caption{\label{tab:cont_WS} As Table \ref{tab:cont_temp} but for the wind speed (WS). }
\begin{tabular}{ccc}
\multicolumn{3}{c}{Wind Speed (WS)}\\
\hline
Param. & Forecast  & Forecast with AR\\
   & the day before  ($\%$) & @ 1h ($\%$) \\
\hline
POD$_{1}$ & 72 & 91 \\
POD$_{2}$ & 48 & 83 \\
POD$_{3}$ & 75 & 93 \\
PC & 65 &  89 \\
EBD & 2 & 0 \\
\hline
\end{tabular}
\end{center}
\end{table}

\begin{table}
\begin{center}
\caption{\label{tab:cont_RH} As Table \ref{tab:cont_temp} but for the relative humidity (RH).}
\begin{tabular}{ccc}
\multicolumn{3}{c}{Relative Humidity (RH)}\\
\hline
Param. & Forecast  & Forecast with AR\\
  & the day before ($\%$)   & @ 1h ($\%$)  \\
\hline
POD$_{1}$ & 91  & 98  \\
POD$_{2}$ & 73 & 95 \\
POD$_{3}$ & 71  & 97  \\
PC & 78 & 97 \\
EBD & 1 & 0 \\
\hline
\end{tabular}
\end{center}
\end{table}

\begin{table}
\begin{center}
\caption{\label{tab:cont_WD} As Table \ref{tab:cont_temp} but for the wind direction (WD) using as a thresholds: 90$^{\circ}$, 180$^{\circ}$ and 270$^{\circ}$.}
\begin{tabular}{ccc}
\multicolumn{3}{c}{Wind direction (WD: 90$^{\circ}$, 180$^{\circ}$ and 270$^{\circ}$)}\\
\hline
Param. & Forecast & Forecast with AR\\
  & the day before ($\%$)    & @ 1h  ($\%$)  \\
\hline
POD$_{1}$ &  78 &  94\\
POD$_{2}$ & 75 &  93\\
POD$_{3}$ & 88  &  94 \\
POD$_{4}$ &  57 &  93 \\
PC & 81 &94 \\
EBD & 2 &  0\\
\hline
\end{tabular}
\end{center}
\end{table}

\begin{table}
\begin{center}
\caption{\label{tab:cont_WD-bis} As Table \ref{tab:cont_temp} but for the wind direction (WD) using as a thresholds: 45$^{\circ}$, 135$^{\circ}$ and 225$^{\circ}$.}
\begin{tabular}{ccc}
\multicolumn{3}{c}{Wind direction (WD: 45$^{\circ}$, 135$^{\circ}$ and 225$^{\circ}$)}\\
\hline
Param. & Forecast & Forecast with AR\\
  & the day before ($\%$)    & @ 1h  ($\%$)  \\
\hline
POD$_{1}$ & 75  & 94 \\
POD$_{2}$ & 62 &  93\\
POD$_{3}$ &  71 &  94 \\
POD$_{4}$ &  84 &   93\\
PC &  74& 94\\
EBD &  1&  0\\
\hline
\end{tabular}
\end{center}
\end{table}

\begin{table}
\begin{center}
\caption{\label{tab:cont_see} As Table \ref{tab:cont_temp} but for the seeing ($\varepsilon$). Values calculated assuming an accuracy of 0.2". We considered the seeing $<$ 1.5". }
\begin{tabular}{ccc}
\multicolumn{3}{c}{Seeing ($\varepsilon$) }\\
\hline
Param. & Forecast  & Forecast with AR\\
  & the day before ($\%$)    & @ 1h  ($\%$)  \\
\hline
POD$_{1}$ & 81 & 99 \\
POD$_{2}$ & 80 & 97 \\
POD$_{3}$ & 65 & 98  \\
PC & 79 & 98 \\
EBD & 14 & 1 \\
\hline
\end{tabular}
\end{center}
\end{table}


\section{AR forecasts in the operational system}
\label{auto}

The study presented in this paper quantifies the improvements obtained in terms of performances of the AR method on short time scales and, in particular, at 1 hour. In this section we describe how this method has been implemented in the ALTA Center, the operational forecast system conceived for the LBTO as we said in the Introduction. We chose to implement the algorithm with N=5 for all the atmospheric parameters and N=3 for the seeing. 
The choice done for the seeing aims to minimise the number of breaks in the sequence of data due to the number of nights in which the telescope dome is closed and we don't have measurements of the seeing. The algorithm has been implemented in the automatic operational system that now works nightly providing forecasts at two different time scales: (a) a forecast of all the classical atmospheric parameters (T, WS, WD and RH and the precipitable water vapour (PWV) and the astroclimatic parameters ($\varepsilon$, $\theta_{0}$, $\tau_{0}$) for the successive night that is available early in the afternoon that we call standard configuration and, (b) a forecast at short time scales. Starting from the sunset, at each full hour (e.g. ..., 02:00 UT, 03:00 UT, 04:00 UT, etc), an AR forecast is calculated and extended for the successive four hours. At each full hour, the forecast extended on four hours is upgraded and shifted of 1 hour as indicated in Fig. \ref{fig:AR_oper}.  

Obviously the AR method can be applied only to all the parameters for which we have in-situ real-time measurements that, at present, are the temperature, RH, WS, WD and seeing. 
So far at Mt.Graham there are no monitors that can provide real-time measurements of the isoplanatic angle ($\theta_{0}$) and the wavefront coherence time ($\tau_{0}$) as well as real-time measurements of the PWV. In the LBTO plans, it is foreseen the implementation in situ of a new generation of Multi Aperture Scintillation Sensor (MASS) that is under development. This should permit to extend the AR forecast at 1h step also to these two parameters ($\theta_{0}$ and $\tau_{0}$) that are extremely important for a set of instruments supported by AO that are running at present such as LUCI\footnote{LUCI is for LBT Utility Camera in the Infrared} with the GLAO\footnote{GLAO is for Ground Layer Adaptive Optics} system ARGOS\footnote{ARGOS is for Advance Rayleigh guided Ground layer adaptive Optics System} \citep{rabien2019} and the Large Binocular Telescope Interferometry (LBTI) \citep{hinz2016} or those that are planned for the near future such as SHARK-VIS\footnote{SHARK is for System for coronography with High order Adaptive optics from R to K bands. Originally a unique instrument, in a successive phase it has been decided to develop two different units in the visible (VIS) and in the near-infrared (NIR).} \citep{pedichini2016}, SHARK-NIR \citep{farinato2018}, iLocator \citep{crepp2016} that will be supported by SOUL\footnote{SOUL is for Single conjugated adaptive Optics Upgrade for the LBT} \citep{pinna2016}, the AO system that will replace FLAO. At the same time, also an instrument providing real-time measurements of the PWV such as LHATPRO \citep{kerber2012} is under evaluation as it should permit an upgrade of the forecasts at short time scale of a parameter such as PWV that is critical for LBTI scientific programs such as those using the nulling interferometry in N band - see HOST project \citep{ertel2018} looking for exozodiacal dust near the habitable zone around nearby, main-sequence stars.

ALTA Center is an operational reality since a couple of years and it is integral part of the operational observing strategy of the LBT \citep{veillet2016} and, since April 2019, it provides forecasts also at short time scale. We have almost completed the implementation of a similar automatic operational system for Cerro Paranal, the site of the Very Large Telescope\footnote{We point out that the operational forecast system for Cerro Paranal is not, at present, an official ESO tool but it is the result of a research study.}. In this astronomical site we can access also to in-situ real-time measurements of $\theta_{0}$, $\tau_{0}$ and the precipitable water vapour (PWV). We expect therefore to be able to achieve forecasts with an equivalent high level in terms of performances for the three principal astroclimatic parameters: seeing, isoplanatic angle and wavefront coherence time.


\section{Conclusions}
\label{concl}

In this paper we analyse for the first time the possibility to provide forecasts of a few fundamental atmospheric parameters (temperature, wind speed, wind direction and relative humidity) as well as astroclimatic parameters such as the seeing at short time scales (order of one hour). This time scale is by far the most critical one for the science operations of top-class telescopes for all those programs using instrumentation supported by AO. The study is applied to Mt. Graham, the site of the Large Binocular Telescope (LBT) where we have an operational forecast system already running nightly, the ALTA Center.
We proposed to use a filtering technique to provide forecasts at short time scale, more precisely we use an auto-regressive technique (AR) based on the simultaneous use of temporal series of real-time measurements performed in situ and predictions provided by a non-hydrostatic atmospheric model Astro-Meso-Nh model.
We demonstrated that the model performances are improved by a not negligible quantities for all the parameters and that a gain is still visible for a few hours. The gain is maximum at one hour and it decreases with the time until it vanishes completely when effect of the knowledge of the in-situ observations at present time loose its positive influence on the future. The values of this thresholds are between 4 and 6 hours, depending on the parameter (Fig.\ref{fig:gain}).

The AR technique for the calculation of forecast at a time scale of one hour produces a gain on model performances of a factor 2.7 up to almost 5 (depending on the atmospheric parameter).
We quantified the performances of the forecast method using different statistical operators. From one side the bias, RMSE and $\sigma$. From the other side, the percentage of correct detection (PC), the probability of detection (POD) and the extremely bad detection (EBD) retrieved from the calculation of the contingency tables. For the time scale of one hour, using the AR filter we obtain: a RMSE=0.25$^{\circ}$ for the temperature, a RMSE=2.91$\%$ for the relative humidity, a RMSE=1ms$^{-1}$ for the wind speed, a RMSE=9.73 degrees for the wind direction when we filter out the wind speed weaker than 3ms$^{-1}$ and a RMSE=0.11" for the seeing. 

Looking at the analysis from the point of view of the contingency tables and probability of detection, we find that all the POD$_{i}$ are well above 90$\%$ reaching in many cases (seeing, temperature, RH and WD) more than 95$\%$. This condition is already very close to the saturation with small space for further improvement. The WS also presents excellent performances for the POD$_{i}$ larger than 90$\%$. POD$_{2}$ is slightly weaker (83$\%$) and tells us that is slightly more difficult to discriminate between the first tertile (5.52 ms$^{-1}$) and the third tertile (9.98 ms$^{-1}$). Besides that, the system is extremely efficient in predicting the weak wind speed (with a RMSE=0.7ms$^{-1}$) and in predicting the very strong wind speed (with a RMSE=1.2ms$^{-1}$), that makes the tool very useful to face the {\it low wind effect} in high contrast imaging instruments (see Section \ref{results}) and to identify the interval of time characterised by very strong wind (WS $>$ 10ms$^{-1}$). 

Results obtained for the optical turbulence, and more precisely with the seeing, are extremely satisfactory. We proved that the AR technique allows us to reach a RMSE of the order of 0.11" at 1 hour and POD$_{i}$ of the order of 98$\%$. The most relevant result obtained in this study is definitely related to the seeing. The most critical POD$_{1}$ i.e. the probability to detect a seeing weaker than the first tertile is equal to 81$\%$ for the standard configuration and it is equal to 98$\%$ for the AR method at 1h time step. This definitely represents a fundamental milestone for the implementation of the flexible-scheduling of ground-based top-class telescopes.  

Besides that, we quantified the gain obtained by the AR approach with respect to the use of pure real-time measurements i.e. the persistence method putting in evidence that the percentage of RMSE gain is between 35\% and 63\% and it is therefore far from being negligible.

Once validated, we implemented this method in the automatic and operational forecast system conceived for the LBTO named ALTA Center. The outputs of such a forecast system are currently automatically injected into the software driving the science operations at the LBTO. At our knowledge, this is the first automatic operational system providing this kind of information, at least in the astronomical context. 

We are implementing a similar automatic and operational system for the VLT and, in this case, we will be able to predict at short time scales, also the isoplanatic angle and the wavefront coherence time thanks to the presence of instrument providing real-time measurements of these parameters. It will be interesting to quantify the performances of this method on different sites. We point out that, at present, this is not an official operational ESO tool.

In terms of filtering techniques, it is our intention to refine our results to evaluate if other methods such as Kalman or machine learning, and multiple ways to use them, might provide supplementary improvements of the technique.

\section*{Acknowledgements}
This work is funded by the contract ENV001 (LBTO). The authors acknowledge the LBTO Director, Christian Veillet and the LBT Board for supporting them in the development of the ALTA Center. The authors thanks the LBTO staff, and in particular Matthieu Bec, for their collaboration in keeping synchronised the real-time LBT telemetry and the ALTA Center. The authors thanks the Meso-Nh users supporter team who constantly works to maintain the model by developping new packages in progressing model versions. Initialisation data come from the GCM of the ECMWF. 






\newpage

%
%
\appendix
\section{Sample used for the statistics analysis}
\label{annex_a}

Table \ref{tab:n_nights} summarises the effective number of nights used for the analysis for each parameter. For temperature and relative humidity we have 351 nights i.e. just a few nights have been missed, for WS and WD we have a total of 324 nights and for the seeing we have a total of 229 nights. The reasons for the missing nights are of different nature. For atmospheric parameter, the reason is mainly given by sporadic failure of the sensors that, for some reason, did not work in a few nights. We note that the anemometers (providing WS and WD measurements) failed for a slightly larger number of nights than temperature and relative humidity. In the case of the seeing the justification for a smaller number of night is mainly due to the fact that measurements are performed with the LBT-DIMM located inside the LBT dome. This means that when the dome is close for whatever reason, measurements are missed. If we consider the shut-down period of LBT (1.5 months in July-August) plus the number of nights lost because of bad weather on 2018 and we subtract to the total number of 365 in one year, we find exactly the number of nights (229) reported in Table \ref{tab:n_nights}, that corresponds to the allocated time of LBT on 2018 ($\sim$ 64$\%$ of the total time).

\begin{table}
\begin{center}
\caption{\label{tab:n_nights} Number or nights used for the analysis of model performances using the AR method. They are extracted from the sample of 365 nights of 2018.}
\begin{tabular}{lccccc}
\hline
   2018 year   & T  & RH  & WS & WD & Seeing \\
\hline
   Num. of nights & 351 & 351& 324 & 324 & 229 \\
\hline
\end{tabular}
\end{center}
\end{table}

%
%

\section{Definitions of: Contingency Tables, PC, POD and EBD}
\label{annex_b}

Table \ref{tab:ct_gen33} is an example of a generic 3$\times$3 contingency table where the observations and simulations are divided into three categories delimited by two thresholds. PC, POD$_{i}$ and EBD can be defined using {\it a},{\it b},{\it c},{\it d},{\it e},{\it f},{\it g},{\it h},{\it i} (number of times in which an observation and a simulation fall inside each category) and N (the total events). The percentage of correct detection PC is defined in Eq.\ref{eq:pc2} where PC = 100\% is the best score; the probability to detect the value of a parameter inside a specific range of values (POD$_{i}$) is given by Eq.\ref{eq:pod1}-Eq.\ref{eq:pod3} where POD$_{i}$ = 100\% is the best score. The extremely bad detection (EBD) probability is given by Eq.\ref{eq:ebd} where EBD = 0\% is the best score. For a total random prediction and in case of a 3$\times$3 contingency table we have {\it a} = {\it b} = ... = {\it i} = {\it N/9} and PC = POD$_{i}$ = 33\% and EBD = 22.2\%.

\begin{equation}
PC=\frac{a+e+i}{N} \times 100; 0\%\leq PC\leq 100\%
\label{eq:pc2}
\end{equation}
\begin{equation}
POD(event_1)=\frac{a}{a+d+g} \times 100; 0\%\leq POD\leq 100\%
\label{eq:pod1}
\end{equation}
\begin{equation}
POD(event_2)=\frac{e}{b+e+h} \times 100; 0\%\leq POD\leq 100\%
\label{eq:pod2}
\end{equation}
\begin{equation}
POD(event_3)=\frac{i}{c+f+i} \times 100; 0\%\leq POD\leq 100\%
\label{eq:pod3}
\end{equation}
\begin{equation}
EBD=\frac{c+g}{N} \times 100; 0\%\leq EBD\leq 100\%
\label{eq:ebd}
\end{equation}

\begin{table}
\begin{center}
\caption{\label{tab:ct_gen33} Generic 3$\times$3 contingency table.}
\begin{adjustbox}{max width=\columnwidth}
\begin{tabular}{cccccc}
\multicolumn{2}{c}{\multirow{2}{*}{Intervals}} & \multicolumn{3}{c}{\bf OBSERVATIONS} &\\
\multicolumn{2}{c}{ } & 1 & 2 & 3 & Total \\
\hline
\multirow{10}{*}{\rotatebox{90}{\bf MODEL}} & & &  &\\
 & \multirow{2}{*}{1}  & a  &  \multirow{2}{*}{b}  &  \multirow{2}{*}{c} &  a+b+c \\
 &           & (hit 1) & & &  1 (Model)\\
 &      &    &    &  &  \\
 & \multirow{2}{*}{2}  & \multirow{2}{*}{d}  & e & \multirow{2}{*}{f} & d+e+f\\
 &           &    & (hit 2) & & 2 (Model)\\
 &      &    &   &  &   \\
 & \multirow{2}{*}{3}  & \multirow{2}{*}{g}  & \multirow{2}{*}{h} & i & g+h+i\\
 &           &    &  & (hit 3) & 3 (Model)\\
 &      &    &   &  &   \\
\hline
 & \multirow{2}{*}{Total} & a+d+g           &  b+e+h        & c+f+i & N=a+b+c+d+e+f+g+h+i \\
 &                        & 1 (OBS) & 2 (OBS) & 3 (OBS) & Total of events \\
\end{tabular}
\end{adjustbox}
\end{center}
\end{table}




\bsp	
\label{lastpage}
\end{document}